\documentclass[aps,prl, twocolumn,showkeys]{revtex4}

\usepackage{graphicx}
\usepackage{amssymb}
\usepackage{amsmath}
\usepackage{colordvi}
\usepackage{color}

\newcommand{\rev}[1]{\textcolor{black}{#1}}

\begin{document}

\title{Enhanced mobility of  quantum droplets in periodic lattices}

\author{Yaroslav V. Kartashov$^1$ and Dmitry A. Zezyulin$^2$}

\affiliation{$^1$Institute of Spectroscopy, Russian Academy of Sciences, Troitsk, Moscow, 108840, Russia\\\medskip
$^2$School of Physics and Engineering, ITMO University, St. Petersburg 197101, Russia}

\keywords{Bose-Bose mixtures, Lee-Huang-Yang corrections, Competing nonlinearity}

\date{\today}

\begin{abstract}

We predict that one- and two-dimensional self-bound quantum droplets, forming in Bose-Einstein condensates in the presence of Lee-Huang-Yang (LHY) quantum corrections to the mean-field energy, may demonstrate exceptional mobility in periodic optical lattices and that they may exhibit considerable displacements across the lattice, remaining dynamically stable, even under weak initial phase kicks imparted to them. Mobility properties of quantum droplets are determined by their internal structure and strongly depend on the number of particles in them. We find that due to the peculiar effect of the  LHY quantum corrections, odd (\rev{i.e}, on-site centered) and even (i.e., inter-site-centered) one-dimensional quantum droplets feature alternating mobility and immobility bands closely corresponding to the regions, where translational perturbation mode is unstable and stable, respectively. This picture becomes even richer in two-dimensional case, where odd-odd, even-odd or even-even quantum-droplets also feature alternating mobility and immobility domains, and where, surprisingly, the droplet may be mobile in one direction, but immobile in the orthogonal direction. We link changes in mobility properties with multiple intersections of energy $E(\mu)$ and norm $N(\mu)$ dependencies for droplets with different internal structure.
\end{abstract}

\maketitle

\textit{Introduction.} Great current interest to quantum droplets (QDs) is motivated by their unusual physical properties stemming from quantum corrections to the meanfield energy that qualitatively affect their existence and stability \cite{Petrov2015, Petrov2016} allowing, in particular, formation of stable multidimensional states even in free space. Quantum droplets were observed and extensively studied in   single-component dipolar Bose gases, where collapse due to long-range attractive dipolar interactions can be compensated by the repulsive Lee-Huang-Yang (LHY) correction \cite{Schmitt2016, FerrierBarbut2016, Wachtler2016, Baillie2016, Chomaz2016, Baillie2018}, and in Bose-Bose mixtures, where the repulsive LHY correction becomes important when the intra-component repulsion is compensated by the inter-component attraction leading to stabilization of not only fundamental \cite{Cabrera2018, Semeghini2018, Cheiney2018, hetero, collisions, Cavicchioli2022}, but also of excited states \cite{Li2018, Kartashov2018, Kartashov2020, Kartashov2019, Tengstrand2019}. Current progress in this rapidly expanding field is summarized in several recent reviews  \cite{review2019,review,review2}.

Remarkably, in the presence of strong confinement along one or two coordinates, the dynamics of one- (1D) and two-dimensional (2D) quantum droplets can be described by the reduced models with different nonlinearities \cite{Petrov2015, Petrov2016}. Therefore, the interplay between such nonlinearities and relatively weak periodic \cite{Morera2020, Dong2020, Zheng2021, Pathak2022, Nie2023, Zhou2019, Zhang2019} and radially periodic \cite{Liu2022, Liu2023, Huang2023} optical lattices may be particularly interesting and unusual, since the properties of lattice solitons strongly depend on the nature and law of nonlinear interactions in the system. Notice that the form of  the beyond mean field corrections may depend on the shape of the confining potential, see for instance \cite{Orso2006} for description of quantum fluctuations in the presence of strong confinement in deep periodic two-dimensional lattices.

The breakup of translational invariance by the lattice is known to severally restrict the mobility of nonlinear excitations that tend to radiate when they travel across the lattice \cite{Braznyi2004, Morsch2006, Bagnato2015, Mihalache2021, Kevrekidis2008} that leads to their eventual trapping around one of the potential minima \cite{Morandotti1999, Vicencio2003, Kartashov2004, Kartashov2009}. This is the case also for Bose-Einstein condensates with local cubic interactions, where compact solitons moving across the lattice are quickly trapped (and only very broad wavepackets show mobility) \cite{Ahufinger2004, Dabrowska2004, Ahufinger2005}. Thus, one of the intriguing open questions, important also for understanding of dynamics of nonlinear waves in various periodic media, is whether nonlinearities specific for QDs can grant them enhanced mobility in optical lattices?

Notice that previously mobility of lattice solitons was studied mainly in distinct physical systems, such as nonlinear optical media, where different degrees of mobility were encountered in discrete models with cubic-quintic \cite{Champneys, Maluchkov2008, Chong2009, Mejia2013} or saturable \cite{Champneys, Hadzievski2004, Melvin2006, Maluchkov2006, Vicencio2006, Oxtoby2007, Melvin2008, Syafwan2012, Naether2011} nonlinearities, in materials with nonlocal nonlinearity \cite{Xu2005}, in competing linear and nonlinear pseudo-potentials \cite{Kartashov2011, Rapti2007, Kartashov2008}, or in models with more complicated nonlocal inter-site nonlinearities \cite{Champneys, Oster2003, Dmitriev2006, Dmitriev2007, Peli2006, Susanto2007}. In most cases, the existence of steadily travelling solitons was related either to vanishing Peierls-Nabarro (PN) barrier (i.e., the energy difference between the onsite- and intersite-centered solitons) or to the onset of dynamical instabilities. In matter-wave systems with LHY quantum contribution such mobility enhancement has never been considered, to the best of our knowledge.

In this paper we show that quantum droplets in optical lattices that exist in stable form due to nontrivial competition between cubic nonlinearity and LHY correction offer very rich opportunities for realization of mobile states that can move across the lattice practically without radiation, provided that certain conditions for norm and symmetry of quantum droplet are fulfilled. In the effectively 1D geometry, we find that continuous families of QDs feature mobility and immobility bands that alternate for odd and even states. We find that mobility properties change around energy intersections for odd and even solutions (which can be heuristically interpreted as a continuous counterpart of the vanishing PN barrier). We report also on the unprecedented mobility of quantum droplets in 2D case, where such states can be mobile in one direction, but immobile in the orthogonal direction. These results are in sharp contrast to previous findings in optical and matter-wave media, where 2D lattice solitons demonstrate very restricted mobility due to strong radiative losses or decay/collapse already after crossing of several of lattice periods, especially in systems with cubic nonlinearity \cite{Vicencio2006,Kartashov2009}.

\begin{figure}
	\begin{center}
		\includegraphics[width=0.99\columnwidth]{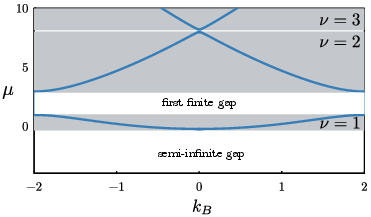}%\\%
	\end{center}
	\caption{\rev{The lowest part of bandgap structure produced by the 1D optical lattice $V(x) = -V_0\cos(2\kappa x)$  with $V_0=2$ and $\kappa=2$. Blue curves show the dependencies $\mu_\nu(k_B)$, $\nu=1,2, 3$, and the gray areas correspond to values of $\mu$ that belong to spectral gaps. The two lowest spectral bands are separated  by the first finite gap. In this paper we consider quantum droplets situated in the semi-infinite gap, i.e., with chemical potentials $\mu$  below the first band with $\nu=1$. For interpretation of the references to color in this figure legend, the reader is referred to the web version of this article.} }
	\label{fig0}
\end{figure}

\textit{Mobility in 1D.} First we consider a quasi-1D Bose-Bose mixture with identical components, which can be modelled by the following dimensionless equation \cite{Petrov2016,AstaMalo}:
\begin{equation}
\label{eq:1D}
i\Psi_t =  -  \frac{1}{2}\Psi_{xx} +  V(x) \Psi -  g|\Psi| \Psi +  \delta  |\Psi|^2 \Psi,
\end{equation}
where positive coefficients $\delta$ and $g$ stand for the effective strengths of the meanfield cubic nonlinearity and the attractive LHY correction, respectively. We use the values of nonlinear coefficients $g=1$ and $\delta =0.4$ that ensure the formation of quantum droplets \cite{Parisi1, Parisi2, Kartashov2022}. Due to the opposite signs, the nonlinear terms are competing, which is a key factor for enhanced mobility. The mixture is loaded in optical lattice $V(x) = -V_0\cos(2\kappa x)$, where $V_0$ and $\kappa$ are the lattice depth and spatial period $\ell = \pi / |\kappa|$, respectively. To ensure validity of Eq. (\ref{eq:1D}), we consider here a relatively shallow lattice with $V_0=2$ and $\kappa=2$, such that in physical units the lattice depth is comparable with the recoil energy (i.e. our regime is far from tight-binding one). Among conserved quantities of system (\ref{eq:1D}) is the number of particles $N = \int_{-\infty}^\infty |\Psi|^2dx$ and energy $E = \int_{-\infty}^\infty (  \frac{1}{2} |\Psi_x|^2 +  V(x) |\Psi|^2  - \frac{2 g}{3} |\Psi|^3  + \frac{\delta}{2}|\Psi|^4)dx$.  {We note that number of particles $N$ is introduced for the dimensionless equation (\ref{eq:1D}) and therefore $N$ can be fractional. The actual number of particles $\mathcal{N}$   scales as $\mathcal{N} \sim 10^3\ldots 10^4 \times N$, depending on chosen physical units.}

\rev{
Let us first briefly address the system in the absence of interactions. This case   formally corresponds to $g = \delta = 0$ in Eq.~(\ref{eq:1D}). The Floquet theory dictates that solutions can be found in the form of Bloch waves $\Psi(x,t) = e^{i[k_Bx -\mu(k_B)t] } u(x; k_B)$, where $k_B$ is the Bloch momentum in the Brillouin zone $k_B \in [-\pi/\ell, \pi/\ell)$, $\mu(k_B)$ is the chemical potential, and functions $u(x; k_B)$ are $\ell$-periodic in $x$. The one-dimensional optical lattice produces a bandgap structure which consists of a countable set of spectral bands with $\mu = \mu_\nu(k_B)$ and  $u = u_\nu(x; k_B)$, where we have introduced index $\nu=1,2, \ldots$ to enumerate the spectral bands. The lowest part of the bandgap structure for our parameters is displayed in Fig.~\ref{fig0}. In what follows, we will consider QDs with chemical potentials $\mu$ situated in the semi-infinite gap, i.e., below the first spectral band.}

\begin{figure}
	\begin{center}
		\includegraphics[width=0.99\columnwidth]{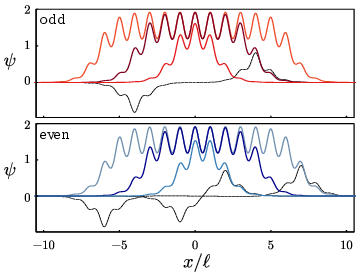}%\\%
	\end{center}
	\caption{Profiles of odd (left panel) and even (right panel) QDs coexisting at $N = 6.6$, $26.0$, and $45.4$. For unstable solutions (odd droplet with $N=26.0$ and even droplets with $N=6.6$ and $45.4$) we also show with black dashed lines the antisymmetric translational perturbation modes.}
	\label{fig1}
\end{figure}

\begin{figure}
	\begin{center}
		\includegraphics[width=0.99\columnwidth]{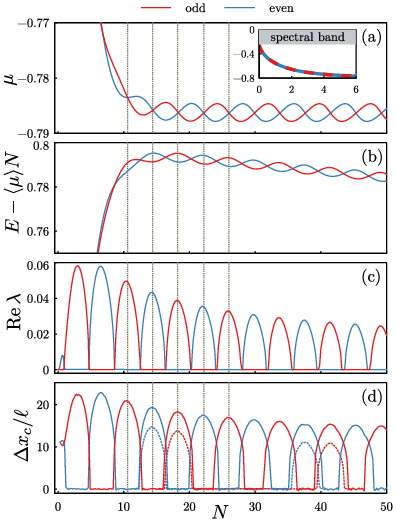}%\\%
	\end{center}
	\caption{(a) Chemical potential $\mu$ vs. number of particles $N$ for odd (red color) and even (blue color) QDs; the inset shows the same dependencies in the region of small $N$, where the families bifurcate from the edge of the spectral band shown with the gray area; in this region the dependencies for odd and even solutions are indistinguishable on the scale of the inset. (b) Normalized energy $E - \langle \mu \rangle N$ (lower panel), where $ \langle \mu \rangle = -0.786$ is a coefficient which is  introduced to remove the irrelevant linear slope. (c) Maximal instability rates. (d) The displacement of the integral center $\Delta x_c = x_c(T)-x_c(0)$ of the droplet plotted in units of the lattice period $\ell$  at $T=2\cdot10^{3}$ vs. number of particles $N$. The initial kick $e^{i\alpha x}$ with $\alpha=0.01$ has been imparted to initially quiescent droplets. Several mobility bands calculated for $\alpha=0.005$ are plotted with dashed curves. Vertical dotted lines in all panels indicate numbers of particles for droplets whose dynamics is shown in Fig.~\ref{fig3} (first three lines) and Fig.~S1  of Supplementary material \cite{SM} (last two lines).   For interpretation of the references to color in this figure legend, the reader is referred to the web version of this article. 
 }
	\label{fig2}
\end{figure}    

Stationary quiescent quantum droplets admit the representation $\Psi(x,t) = e^{-i \mu   t}\psi(x)$, where $\mu$ is the chemical potential and $\psi(x)\to 0$ as $x\to\pm\infty$. It is known  \cite{Louis,Peli,Yang} that in a system with purely meanfield repulsive nonlinearity solitons do not appear in the semi-infinite forbidden gap of the spectrum. However, the attractive LHY correction enables the existence of QDs in the semi-infinite gap. We focus on two families of solutions that can be denoted as odd and even QDs. This terminology reflects the spatial shape of corresponding wavefunctions $\psi(x)$, which, respectively, feature  odd or even numbers of local maxima. Odd QDs are centered at the lattice ``sites'' [one of minima of $V(x)$], while even QDs are centered between lattice ``sites'' [i.e. they have at least two equal maxima of $\psi$ on neighboring lattice minima]. Several examples of QDs with different numbers of peaks are shown in Fig.~\ref{fig1}.   Stationary solutions presented in  Fig.~\ref{fig1} and below  have been computed using the iterative Newton method applied to the corresponding  finite-difference problem on a sufficiently large computational window subject to the zero boundary conditions.  The families of odd and even QDs detach from the lowest edge of the spectral band [see the inset in Fig.~\ref{fig2}(a)].  As number of particles $N$ increases, the impact of the repulsive cubic nonlinearity builds up   that leads to the appearance of the lower bound of the chemical potential [Fig. \ref{fig2}(a)]. Remarkably, for sufficiently large $N$ the competition between cubic nonlinearity and LHY correction causes out-of-phase oscillations of chemical potentials $\mu(N)$ for odd and even QDs, leading to multiple intersections of corresponding $\mu(N)$ curves. These oscillations are accompanied by the gradual broadening of the droplets, so that they eventually acquire flat-top shapes as in uniform case \cite{AstaMalo}, but with density modulations due to presence of the lattice. Similar oscillations are observed also in $E(N)$ dependencies for odd and even droplets in Fig.~\ref{fig2}(b).  Notice that the difference of energies $E$ of two different solutions at fixed number of particles $N$ is frequently associated with so-called Peierls-Nabarro barrier that the solution should overcome, when it moves across the lattice (thereby passing between odd and even configurations upon motion). Hence, the points where the difference of energies vanishes correspond to the points, where this barrier vanishes, and hence even minimal kick is sufficient to set droplet in motion. 

Multiple intersections between energies of odd and even QDs indicate on possible recurring exchange of stability between these states. To analyse their stability we substitute the perturbed QD profile $\Psi = e^{-i\mu t}[\psi(x) + \chi_1(x,t) + i\chi_2(x,t)]$, where $\chi_{1,2} = e^{\lambda t}\varphi_{1,2}(x)$ are components of a small perturbation, into Eq.~(\ref{eq:1D}), linearize it around $\psi$, and obtain  a pair or equations:  $\lambda \varphi_1 = -L^- \varphi_2$ and  $\lambda \varphi_2 =  L^+ \varphi_1$, where     $L^\pm = \frac{1}{2} \partial_x^2 + \mu - V(x)  + \frac{g}{2}(3\pm 1)|\psi|  - \delta (2\pm 1) \psi^2$. The instability increment is given by the real part of $\lambda$. Solving this eigenvalue problem for $\lambda$, we obtained a sequence of exactly alternating domains of stability and instability for odd and even QDs, presented in Fig.~\ref{fig2}(c) [change of stability properties occurs practically in the intersections of $E(N)$ curves]. For sufficiently large number of particles, the intervals where QDs are stable (unstable) almost perfectly coincide with the intervals where $d\mu/dN >0\, (<0)$. Thus, in their stability regions odd and even droplets formally obey the ``anti-Vakhitov-Kolokolov'' stability criterion \cite{VK}. \rev{In the case at hand, the violation of ``standard'' Vakhitov-Kolokolov (VK) criterion results from the presence of the optical lattice and from the fact that our system is characterized by competition between attractive and repulsive nonlinearities that may considerably change the perturbation spectrum. In earlier literature, similar examples of successful usage of the anti-VK stability criterion have been reported for QDs in 2D optical lattices \cite{Zheng2021}, for spinor 2D QDs with Rashba-type spin-orbit coupling \cite{LiMalomed}, and  for discrete QDs \cite{Discrete1,Discrete2}.} 

Alternating stability/instability domains are associated with a  translational mode composed of a pair of antisymmetric eigenfunctions $\varphi_{1,2}(x) = -\varphi_{1,2}(-x)$. In a uniform medium, this  translational mode would  correspond to an identically zero eigenvalue $\lambda=0$. However, as the lattice is introduced, the zero eigenvalue gives birth to  a pair of opposite eigenvalues  which repeatedly transform from being purely imaginary (within stability regions) to purely real (within instability regions) as the number of atoms $N$ increases. Oscillations of the translational eigenvalues for odd and even QDs are out-of-phase and hence result in the alternating stability bands in Fig.~\ref{fig2}(c).  Notice that when such unstable translational mode is added as a perturbation to initial state, its development at the initial stages of evolution is manifested namely as a transverse shift of corresponding droplet. 

\begin{figure}%[t!]
	\begin{center}
		\includegraphics[width=0.99\columnwidth]{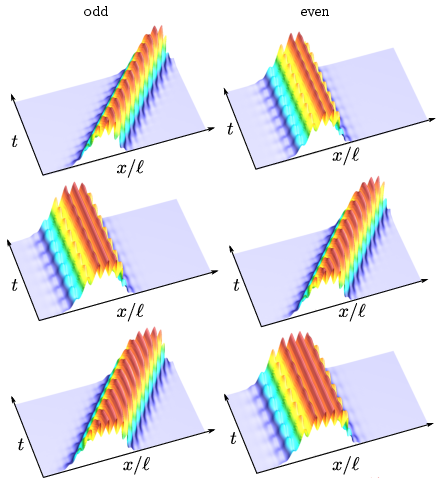}%\\%
	\end{center}
	\caption{Evolution dynamics of odd (\rev{left} column) and even (\rev{right}  column) droplets coexisting at $N=10.6, 14.4, 18.2$; the corresponding values of $N$ are highlighted with vertical dotted lines in Fig.~\ref{fig2}. For each droplet, an initial phase kick $e^{i\alpha  x}$ has been imparted, with $\alpha=0.01$. In all panels the spatial axis $x/\ell$ occupies 21 spatial periods, and the time axis   increases from $t=0$ to $t=10^3$.}
	\label{fig3}
\end{figure}

These stability properties of QDs find their manifestation in dramatically enhanced mobility in certain intervals of $N$ --- i.e. steady drifts of initially kicked states practically without radiative losses. In Fig.~\ref{fig2}(d) we plot the displacement of the integral center of the droplet $\Delta x_c$ versus $N$ [here the integral center position is defined as $x_c(t)  =  N^{-1}\int_{-\infty}^\infty x |\Psi(x,t)|^2dx$], stimulated by the initial kick $\sim e^{i\alpha x}$ with small $\alpha$, after sufficiently large evolution time $T$.  One observes alternating bands with sharp boundaries between mobile and immobile droplets with a given symmetry (odd or even), which closely correlate with corresponding stability and instability domains. Figure~\ref{fig3} shows dynamics of kicked odd and even 1D QDs with the same number of particles $N$ corresponding to the consecutive intersections of $\mu(N)$ curves in Fig. \ref{fig2}(a), clearly highlighting alternating mobility properties. Notice that enhanced mobility can be observed already for compact states, far from the flat-top limit. Dynamical simulations presented in Fig.~\ref{fig3} have been performed using the well-known pseudospectral split-step method \cite{Yang}. Examples of dynamical mobility and immobility of flat-top states are presented in the  Supplementary material \cite{SM}.

The results presented above have been obtained for a relatively shallow lattice. While the enhanced mobility remains robust under a moderated increase of the lattice depth,  our results indicate that a very deep lattice   suppresses the mobility. For instance,  for $V_0=15$ mobile droplets have been found  only in a small vicinity of the point, where stability of odd solitons changes for the first time, and only for the odd solution. 

\rev{In possible experiments with mixtures of two $^{39}$K Bose-Einstein condensates in two different hyperfine states \cite{Cabrera2018,Semeghini2018}, condensation in odd or even droplets can be achieved by using a combination of an anisotropic harmonic trap and periodic optical lattice that can be shifted with respect to the trap to produce either local minimum (for odd states) or maximum (for even states) of potential at $x=0$. In such experiments the transition into the regime, where droplets may form, typically occurs upon variation of the intra(inter)-species scattering lengths via Feshbach resonance in the external magnetic field, see for details \cite{Cabrera2018,Semeghini2018}. When the formation of a desired state is achieved, harmonic trap can be switched off, while well-developed phase-imprinting methods \cite{Burger1999,Denschlag2000} can be used to create initial kick for the droplet.}

\begin{figure}%[t!]
	\begin{center}
		\includegraphics[width=0.99\columnwidth]{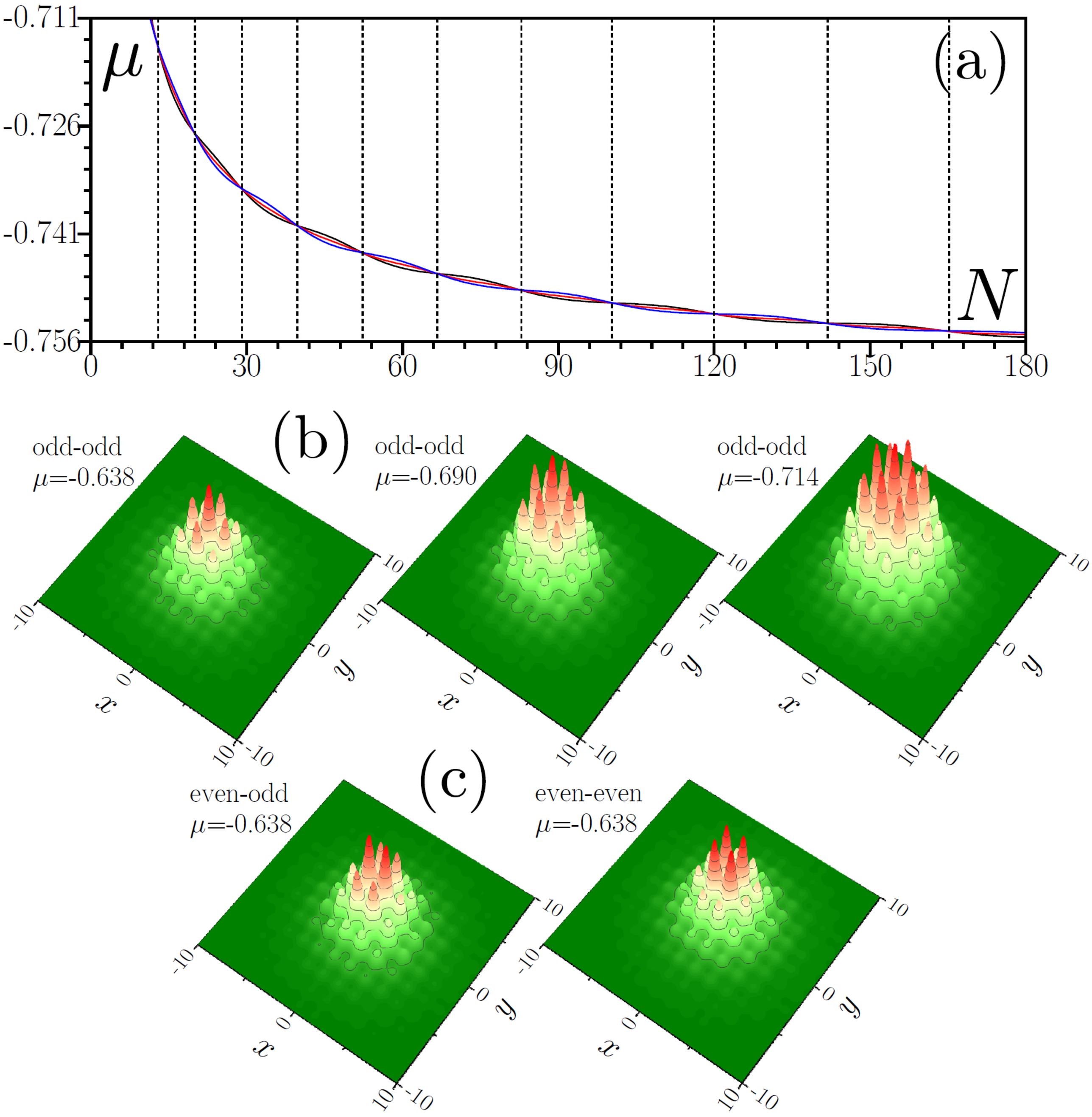}%\\%
	\end{center}
	\caption{(a) Chemical potential versus number of particles in 2D QDs of odd‐odd (black curves), even‐odd (red curves), and even‐even (blue curves) types. Dashed lines highlight intersection points of all three dependencies. (b) Profiles of odd‐odd QDs in consecutive intersection points. (c) Profiles of even‐odd and even‐even QDs in one of the intersection points. Chemical potentials are indicated near corresponding profiles. In all cases $\kappa=2$. For interpretation of the references to color in this figure legend, the reader is referred to the web version of this article.} 
	\label{fig4}
\end{figure}

\begin{figure}[ht!]
	\begin{center}
		\includegraphics[width=0.99\columnwidth]{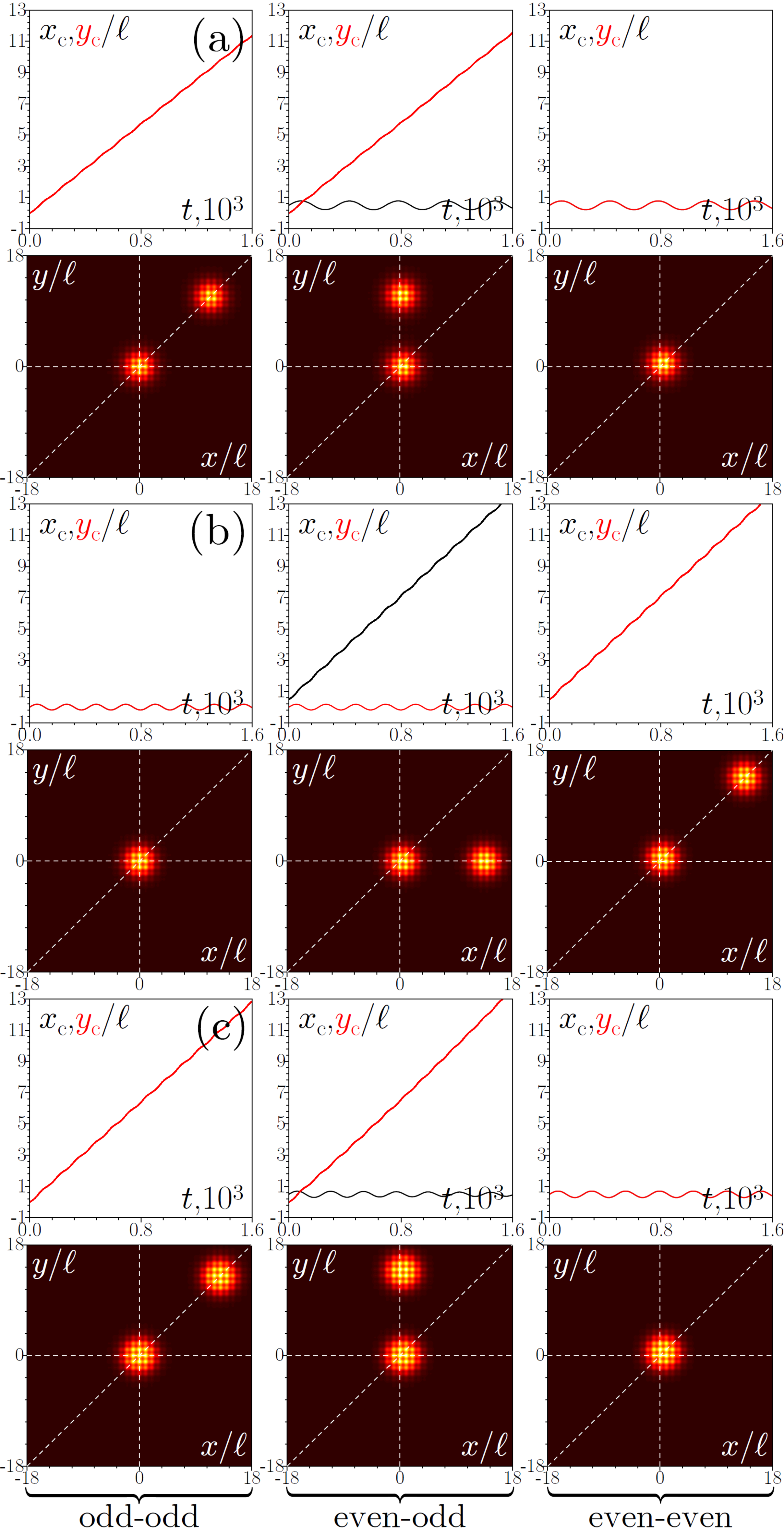}%\\%
	\end{center}
	\caption{Comparison of mobility of three different types of quantum droplets with
chemical potentials $\mu=-0.638$ (a), $\mu=-0.690$ (b), and $\mu=-0.714$ (c) corresponding to consequent intersections of their $\mu(N)$ curves. In each case top row shows coordinates $(x_c, y_c)$  of the integral center of the droplet vs. time, while bottom row shows superimposed snapshots of $|\Psi|$ at $t =0$ and $t =1800$. Dashed white lines are guides for the eye. In all cases the initial kick was imparted on droplet by multiplying its wavefunction with $e^{i \alpha x  + i\alpha y}$ with $\alpha=0.01$. For interpretation of the references to color
in this figure legend, the reader is referred to the web version
of this article.} 
	\label{fig5}
\end{figure}

\textit{Mobility in 2D.} The evolution of the effectively 2D QDs under assumption of identical components in Bose-Bose mixture is governed by the equation \cite{Petrov2016}:
\begin{equation}
\label{eq:2D}
i\Psi_t =  -  \frac{1}{2}(\Psi_{xx} + \Psi_{yy}) +  V(x,y) \Psi +  2 |\Psi|^2  \Psi \ln (2 |\Psi|^2),
\end{equation}
where we now consider two-dimensional lattice $V(x,y)=-2\cos(2\kappa x) - 2\cos(2\kappa y)$ with $\kappa=2$.
%, to be able to compare the results with 1D case. 
Logarithmic nonlinearity %considered here 
describes competition between attraction in the regions with small $|\Psi|^2$ and repulsion in domains, where $|\Psi|^2$ is large enough. The peculiarity of the 2D case is that now the droplet can have the same or different symmetries along the $x$ and $y$ axes that allows to obtain three different droplet families, odd-odd, even-odd, and even-even ones.
%, according to the location of their centers with respect to the lattice sites. 
Corresponding $\mu(N)$ dependencies and representative examples of stationary solutions are shown in Fig.~\ref{fig4}. As in the 1D case, the droplets appear in semi-infinite gap, and their norm rapidly increases as $\mu$ approaches certain minimal value. Remarkably, all \textit{three} families intersect practically at the same number of particles $N=\iint |\Psi|^2dxdy$ indicated in Fig.~\ref{fig4}(a) by the vertical dashed lines. Each subsequent intersection corresponds to addition of humps (broadening) to QD of a given type [Fig. \ref{fig4}(b)]. In contrast to the 1D case, $N$ always increases with decrease of $\mu$. Similar intersections are observed in $E(\mu)$ dependencies (not shown). By analogy with 1D case, one can expect \textit{alternation} of mobility properties for states in different intersections of $\mu(N)$ curves. We thus simulated the dynamics of QDs corresponding to consecutive intersections in the presence of initial phase kick imparted by the multiplication of the input by the $e^{i \alpha x  + i\alpha y}$ factor (thus, we simultaneously test mobility properties in two orthogonal directions). We found unprecedented mobility enhancement for QDs with certain symmetries in this system manifested in their motion over multiple lattice periods practically without radiation (in contrast to previous results on mobility of 2D lattice solitons that tend to quickly decay or get trapped after initial kick), see Fig. \ref{fig5}. For each intersection we observe that one of the states featuring the same symmetry along two axes (i.e., either odd-odd or even-even) is mobile and travels in the diagonal direction, while another one remains nearly quiescent and only weakly oscillates around its initial position. In the next intersection, the former state becomes immobile, while the latter starts moving [compare first and third columns in Fig. \ref{fig5}(a) and (b)]. Remarkably, ``mixed'' (i.e. even-odd) states always feature different mobility properties along the two directions (Fig. \ref{fig5}, middle column), i.e., they start moving either along $x$ or along $y$ axes, and this direction alternates for consecutive intersections. To stress that this alternation of mobility properties persists in next intersections, even for broad flat-top modes, we present their dynamics in the Supplementary material  \cite{SM}.

\paragraph{Conclusion.} Summarizing, we have predicted that 1D and 2D QDs in optical lattices may demonstrate strongly enhanced mobility that depends on the number of particles and internal symmetry of corresponding droplet states. Mobility enhancement is observed even for sufficiently compact QDs, occupying only several lattice sites. The mobility enhancement can be linked to the excitation  of unstable translational perturbation modes. Our results may open the way for the experimental realization of highly mobile self-sustained states in BECs trapped in lattices, and suggest new strategies for control of dynamics of matter waves. 

%\textit{To conclude,} we have uncovered the general patterns that govern the enhanced mobility of one- and two-dimensional quantum droplets in periodic potentials.  

%We also note that the results presented above have been obtained for a relatively shallow lattice. The increase of the lattice depth leads to the suppression of the mobility. Say for $V_0=15$ mobile droplets can be found only in a small vicinity of the first stability change (and only for odd solutions).

\section*{Declaration of competing interest}

The authors declare that they have no known competing financial interests or personal relationships that could have appeared to
influence the work reported in this paper.

\section*{Funding}

The work of Y.V.K. was supported by the research project FFUU-2024-0003 of the Institute of Spectroscopy of the Russian Academy of Sciences. The work of D.A.Z. was supported by the Priority 2030 Federal Academic Leadership Program.

\section*{Data availability}

Data will be made available on request.


\begin{thebibliography}{00}


\bibitem{Petrov2015} D. S. Petrov, Quantum Mechanical Stabilization of a Collapsing Bose-Bose Mixture, Phys. Rev. Lett. {\bf  115}, 155302 (2015).

\bibitem{Petrov2016} D. S. Petrov and G. E. Astrakharchik, Ultradilute Low-Dimensional Liquids, Phys. Rev. Lett. {\bf 117}, 100401 (2016).

\bibitem{Schmitt2016} M. Schmitt, M. Wenzel, F. Böttcher, I. Ferrier-Barbut, and T. Pfau, Self-bound droplets of a dilute magnetic quantum liquid, Nature \textbf{539}, 259 (2016).

\bibitem{FerrierBarbut2016} I. Ferrier-Barbut, H. Kadau, M. Schmitt, M. Wenzel, and T. Pfau, Observation of quantum droplets in a strongly dipolar Bose gas, Phys. Rev. Lett. \textbf{116}, 215301 (2016).

\bibitem{Wachtler2016} F. Wächtler and L. Santos, Ground-state properties and elementary excitations of quantum droplets in dipolar Bose-Einstein condensates, Phys. Rev. A \textbf{94}, 043618 (2016).

\bibitem{Baillie2016} D. Baillie, R. M. Wilson, R. N. Bisset, and P. B. Blakie, Self-bound dipolar droplet: A localized matter wave in free space, Phys. Rev. A \textbf{94}, 021602(R) (2016).

\bibitem{Chomaz2016} L. Chomaz, S. Baier, D. Petter, M. J. Mark, F. Wächtler, L. Santos, F. Ferlaino, Quantum-fluctuation-driven crossover from a dilute Bose-Einstein condensate to a macrodroplet in a dipolar quantum fluid, Phys. Rev. X \textbf{6}, 041039 (2016).

\bibitem{Baillie2018} D. Baillie and P. B. Blakie, Droplet crystal ground states of a dipolar Bose gas, Phys. Rev. Lett. \textbf{121}, 195301 (2018).

%\bibitem{Kadau2016} H. Kadau, M. Schmitt, M. Wenzel, C. Wink, T. Maier, I. Ferrier-Barbut, and T. Pfau, Observing the Rosensweig instability of a quantum ferrofluid,  Nature (London) {\bf  530}, 194 (2016).

%\bibitem{Ferrier-Barbut2016}  I. Ferrier-Barbut, H. Kadau, M. Schmitt, M. Wenzel, and T. Pfau, Observing the Rosensweig instability of a quantum ferrofluid, Phys. Rev. Lett. {\bf  116}, 215301 (2016).

\bibitem{Cabrera2018} C. R. Cabrera, L. Tanzi, J. Sanz, B. Naylor, P. Thomas, P. Cheiney, and L. Tarruell, Quantum liquid droplets in a mixture of Bose-Einstein condensates, Science {\bf 359}, 301 (2018).

\bibitem{Semeghini2018} G. Semeghini, G. Ferioli, L. Masi, C. Mazzinghi, L. Wolswijk,
F. Minardi, M. Modugno, G. Modugno, M. Inguscio, and M.
Fattori, Self-Bound Quantum Droplets of Atomic Mixtures in Free Space, Phys. Rev. Lett. {\bf  120}, 235301 (2018).

\bibitem{Cheiney2018} P. Cheiney, C. R. Cabrera, J. Sanz, B. Naylor, L. Tanzi, and L. Tarruel, Bright Soliton to Quantum Droplet Transition in a Mixture of Bose-Einstein Condensates, Phys. Rev. Lett. {\bf  120}, 135301 (2018).

\bibitem{hetero} C.  D'Errico, A.  Burchianti, M.  Prevedelli, L. Salasnich, F.  Ancilotto, M.  Modugno, F. Minardi, and C. Fort, Observation of quantum droplets in a heteronuclear bosonic mixture, Phys. Rev. Research {\bf  1},  033155 (2019).

\bibitem{collisions} G. Ferioli, G.  Semeghini, L.  Masi, G. Giusti, G. Modugno, M.  Inguscio, A. Gallem\'{\i}, A. Recati,   and M. Fattori,  Collisions of self-bound quantum droplets, Phys. Rev. Lett. {\bf  122},  090401 (2019).

\bibitem{Cavicchioli2022} L. Cavicchioli, C. Fort, M. Modugno, F. Minardi, and A. Burchianti, Dipole dynamics of an interacting bosonic mixture, Phys. Rev. Research \textbf{4}, 043068 (2022).

\bibitem{Li2018} Y. Li, Z. Chen, Z. Luo, C. Huang, H. Tan, W. Pang, and B. A. Malomed, Two-dimensional vortex quantum droplets, Phys. Rev. A \textbf{98}, 063602 (2018).

\bibitem{Kartashov2018} Y. V. Kartashov, B. A. Malomed, L. Tarruell, and L. Torner, Three-dimensional droplets of swirling superfluids, Phys. Rev. A \textbf{98}, 013612 (2018).

\bibitem{Kartashov2020} Y. V. Kartashov, B. A. Malomed, L. Torner, Structured hetero-symmetric quantum droplets, Phys. Rev. Research \textbf{2}, 033522 (2020).

\bibitem{Kartashov2019} Y. V. Kartashov, B. A. Malomed, and L. Torner, Metastability of quantum droplet clusters, Phys. Rev. Lett. \textbf{122}, 193902 (2019).

\bibitem{Tengstrand2019} M. N. Tengstrand, P. Stürmer, E. Ö. Karabulut, and S. M. Reimann, “Rotating binary Bose-Einstein condensates and vortex clusters in quantum droplets,” Phys. Rev. Lett. \textbf{123}, 160405 (2019).

\bibitem{review2019} Y. V. Kartashov, G. E. Astrakharchik, B. A. Malomed, and L. Torner, Frontiers in multidimensional self-trapping of nonlinear fields and matter, Nature Rev. Phys.  {\bf  1},  185  (2019).

\bibitem{review} Z.-H. Luo, W. Pang, B. Liu, Y.-Y. Li, and B. A. Malomed, A new form of liquid matter: Quantum droplets, Front. Phys. {\bf 16}, 32201 (2021).

\bibitem{review2} F.  B\"ottcher, J.-N. Schmidt, J.  Hertkorn, K. S. N. Ng, S. D.  Graham, M. Guo, T.  Langen, and T.  Pfau, New states of matter with 
fine-tuned interactions: quantum droplets and dipolar supersolids, Rep. Prog. Phys. {\bf  84}, 012403 (2021).

\bibitem{Zhou2019}  Z. Zhou, X. Yu, Y. Zou, and H. Zhong, Dynamics of quantum droplets in a one-dimensional optical lattice, Commun. Nonlinear Sci. Numer. Simulat. {\bf 78}, 104881 (2019).

\bibitem{Dong2020} L. Dong,  W. Qi, P. Peng, L. Wang, H. Zhou, C. Huang, Multi-stable quantum droplets in optical lattices, Nonlinear Dyn.  {\bf 102}, 303--310 (2020).

\bibitem{Morera2020}  I. Morera, G.   E. Astrakharchik,  A. Polls, and Bruno Jul\'{\i}a-D\'{\i}az, Quantum droplets of bosonic mixtures in a one-dimensional optical lattice,  Phys. Rev. Research {\bf  2}, 022008(R) (2020).

\bibitem{Zhang2019} X. L. Zhang, X. X. Xu, Y. Y. Zheng, Z. P. Chen, B. Liu, C. Q. Huang, B. A. Malomed, Y. Y. Li,  Semidiscrete quantum droplets and vortices,  Phys. Rev. Lett. {\bf 123}, 133901 (2019).

%\bibitem{Dong2022} Liangwei Dong,  Dongshuai Liu, Zhijing Du, Kai Shi, and Wei Qi, Bistable multipole quantum droplets in binary Bose-Einstein condensates, Phys. Rev. A {\bf  105}, 033321 (2022).

%\bibitem{Zezyulin2023} D. A. Zezyulin, Quasi-one-dimensional harmonically trapped quantum droplets,  Phys. Rev. A {\bf  107}, 043307 (2023).

\bibitem{Zheng2021} Y.-Y. Zheng, S.-T. Chen, Z.-P. Huang, S.-X. Dai, B. Liu, Y.-Y. Li, S.-R. Wang, Quantum droplets in two-dimensional optical lattices, Front. Phys.
{\bf 16}, 22501 (2021).

\bibitem{Pathak2022} M. R. Pathak and  A. Nath, Droplet to soliton crossover at negative temperature in presence of bi‑periodic optical lattices, Sci. Rep. {\bf 12},  18248 (2022).

\bibitem{Nie2023}  Y. Nie, J.-H. Zheng,  and T. Yang, Spectra and dynamics of quantum droplets in an optical lattice, Phys. Rev. A {\bf 108}, 053310 (2023).

\bibitem{Liu2022}  B. Liu, Y.-X. Chen, A.-W. Yang, X.-Y. Cai, Y. Liu, Z.-H. Luo, X.-Z. Qin, X.-D. Jiang, Y.-Y. Li, and B. A Malomed, Stable quantum droplets with higher-order vortex in radial lattices,  New J. Phys. {\bf 24},  123026   (2022).

\bibitem{Liu2023} B. Liu, X. Cai, X. Qin, X. Jiang, J. Xie, B. A. Malomed, and Y. Li, Ring-shaped quantum droplets with hidden vorticity in a radially periodic potential, Phys. Rev. E {\bf 108}, 044210 (2023).

\bibitem{Huang2023}
H. Huang, H.-C. Wang, G. Chen, M. Chen, C. S. Lim, K.-C. Wong, Stable quantum droplets with higher-order vortex in radial lattices, Chaos, Solitons and Fractals {\bf 168},  113137  (2023).

\bibitem{Orso2006}
G. Orso, C. Menotti, and S. Stringari, "Quantum fluctuations and Collective Oscillations of a Bose-Einstein Condensate in a 2D Optical Lattice," Phys. Rev. Lett. \textbf{97}, 190408 (2006).

\bibitem{Braznyi2004} V. A. Braznyi and V. V. Konotop, Theory of Nonlinear Matter Waves in Optical Lattices, Mod. Phys.  Lett.  B {\bf  18},  627  (2004).

\bibitem{Morsch2006} O. Morsch and M. Oberthaler, Dynamics of Bose-Einstein condensates in optical lattices, Rev. Mod. Phys. {\bf 78}, 179 (2006).

\bibitem{Kevrekidis2008} \textit{Emergent Nonlinear Phenomena in Bose-Einstein Condensates}, edited by P. G. Kevrekidis, D. J. Frantzeskakis, and R.
Carretero-González, Springer Series on Atomic, Optical,
and Plasma Physics Vol. 45 (Springer, New York, 2008).

\bibitem{Bagnato2015} V. S. Bagnato,  D. J.   Frantzeskakis, P. G.   Kevrekidis, B. A.   Malomed,  D.  Mihalache,   Bose-Einstein Condensation: Twenty Years After, 
Rom. Rep.  Phys.  {\bf 67}, 5  (2015).

\bibitem{Mihalache2021} D. Mihalache, Localized structures in optical and matter-wave media: A selection of recent studies, Rom. Rep. Phys. {\bf 73}, 117 (2021).

\bibitem{Morandotti1999} R. Morandotti, U. Peschel, J. S. Aitchison, H. S. Eisenberg, and Y. Silberberg, Dynamics of Discrete Solitons in Optical Waveguide Arrays, Phys. Rev. Lett. \textbf{83}, 2726 (1999).

\bibitem{Vicencio2003} R. A. Vicencio, M. I. Molina, and Y. S. Kivshar, Controlled switching of discrete solitons in waveguide arrays, Opt. Lett. \textbf{28}, 1942 (2003).

\bibitem{Kartashov2004} Y. V. Kartashov, A. S. Zelenina, L. Torner, and V. A. Vysloukh, Spatial soliton switching in quasi-continuous optical arrays, Opt. Lett. \textbf{29}, 766-768 (2004).

\bibitem{Kartashov2009} Y.  V. Kartashov, V. A. Vysloukh, and L. Torner,  Soliton Shape and Mobility Control in Optical Lattices, Progress in Optics {\bf   52}, 63 (2009).

%\bibitem{Dong2022} Liangwei Dong, Kai Shi, and Changming Huang,  Internal modes of two-dimensional quantum droplets, Phys. Rev. A {\bf 106}, 053303 (2022).

%\bibitem{Flach} S. Flach and  A. V. Gorbach, Discrete breathers --- Advances in theory and applications, Phys. Rep. {\bf   467} 1--116 (2008).

\bibitem{Ahufinger2004}
V. Ahufinger,  A. Sanpera,  P. Pedri,  L. Santos,  and M. Lewenstein,  Creation and mobility of discrete solitons in Bose-Einstein condensates, Phys. Rev. A {\bf 69}, 053604 (2004).

\bibitem{Dabrowska2004} B. J. Dabrowska, E. A. Ostrovskaya, and Yu.  S. Kivshar,  Interaction of matter-wave gap solitons in optical lattices, J. Opt. B: Quantum Semiclass. Opt. {\bf 6}, 423 (2004).

\bibitem{Ahufinger2005} V. Ahufinger and A. Sanpera, Lattice Solitons in Quasicondensates, Phys. Rev. Lett.  {\bf 94}, 130403 (2005).

\bibitem{Maluchkov2008} A. Maluckov, Lj. Hadzievski, and B. A. Malomed, Staggered and moving localized modes in dynamical lattices with the cubic-quintic nonlinearity, Phys. Rev. E {\bf 77},  036604  (2008).

\bibitem{Chong2009}
C. Chong,  R. Carretero-Gonz\'alez, B. A. Malomed, P. G. Kevrekidis, Multistable solitons in higher-dimensional cubic–quintic nonlinear Schr\"odinger lattices, Physica D  {\bf 238},  126--136  (2009).

\bibitem{Mejia2013}
C. Mej\'{\i}a-Cort\'es, R. A. Vicencio, and B.  A. Malomed, Mobility of solitons in one-dimensional lattices with the cubic-quintic nonlinearity, Phys. Rev. E {\bf 88}, 052901 (2013).

\bibitem{Champneys} A. R. Champneys, V. M. Rothos, and T. R.O. Melvin, Traveling Solitary Waves in DNLS Equations, in: P. G. Kevrekidis, \textit{ The Discrete Nonlinear Schr\"odinger Equation} (Springer, Berlin, Heidelberg, 2009).

\bibitem{Melvin2006} T. R. O. Melvin,  A. R. Champneys,  P. G. Kevrekidis,  and J. Cuevas, Radiationless Traveling Waves in Saturable Nonlinear Schr\"odinger Lattices, Phys. Rev. Lett. {\bf 97}, 124101 (2006).

\bibitem{Oxtoby2007} O. F. Oxtoby and I. V. Barashenkov, Moving solitons in the discrete nonlinear Schr\"odinger equation, Phys. Rev. E {\bf 76}, 036603 (2007).

\bibitem{Naether2011} U.  Naether,  R. A. Vicencio, M. Stepi\'c, Mobility of high-power solitons in saturable nonlinear photonic lattices, Opt. Lett. {\bf 36},  1467-1469 (2011).

%\bibitem{Alfimov2019} G. L. Alfimov, A. S. Korobeinikov, C. Lustri, and D. E. Pelinovsky, Nonlinearity  {\bf 32}, 3445 (2019).

\bibitem{Vicencio2006} R. A. Vicencio and M. Johansson, Discrete soliton mobility in two-dimensional waveguide arrays with saturable nonlinearity,  Phys. Rev. E. {\bf 73},  046602 (2006).

\bibitem{Melvin2008} T. R. O. Melvin, A. R. Champneys,  P. G. Kevrekidis, and J. Cuevas, Travelling solitary waves in the discrete Schr\"odinger equation with saturable  nonlinearity: Existence, stability and dynamics, Physica D  {\bf 237},  551--567  (2008).

\bibitem{Hadzievski2004} L. Hadzievski, A. Maluckov, M. Stepi\'c, and D. Kip, Power Controlled Soliton Stability and Steering in Lattices with Saturable Nonlinearity, Phys. Rev. Lett. {\bf 93}, 033901 (2004).

\bibitem{Maluchkov2006} A. Maluckov, L. Hadzievski, M. Stepi\'c,  Bifurcation analysis of the localized modes dynamics in lattices with saturable nonlinearity, Physica D {\bf 216},  95--102  (2006).

\bibitem{Syafwan2012} M. Syafwan,  H. Susanto, S. M. Cox, and B. A. Malomed, Variational approximations for traveling solitons in a discrete nonlinear Schr\"odinger equation, J. Phys. A: Math. Theor. {\bf 45}, 075207  (2012).

\bibitem{Xu2005} Z. Xu, Y. V. Kartashov, and L. Torner, Soliton Mobility in Nonlocal Optical Lattices, Phys. Rev. Lett. \textbf{95}, 113901 (2005).

\bibitem{Kartashov2011} Y. V. Kartashov, B. A. Malomed, and L. Torner, Solitons in nonlinear lattices, Rev. Mod. Phys. {\bf 83}, 247 (2011).

\bibitem{Rapti2007} Z. Rapti, P. G. Kevrekidis, V. V. Konotop, and C. K. R.
T. Jones, Solitary Waves Under the Competition of Linear and Nonlinear Periodic Potentials, J. Phys. A \textbf{40}, 14151 (2007).

\bibitem{Kartashov2008} Y. V. Kartashov, V. A. Vysloukh, and L. Torner, Soliton modes, stability, and drift in optical lattices with spatially modulated nonlinearity, Opt. Lett. \textbf{33}, 1747-1749 (2008).

\bibitem{Oster2003} M. \"Oster, M. Johansson, and A. Eriksson, Enhanced mobility of strongly localized modes in waveguide arrays by inversion of stability, Phys. Rev. E {\bf 67}, 056606 (2003).

\bibitem{Dmitriev2006} S. V. Dmitriev, P. G. Kevrekidis, A. A. Sukhorukov, N. Yoshikawa, S. Takeno,  Discrete nonlinear Schr\"odinger equations free of the Peierls–Nabarro potential, Phys. Lett. A {\bf   356},  324-332  (2006).

\bibitem{Peli2006} D. E. Pelinovsky, Translationally invariant discrete nonlinear Schr\"odinger lattices, Nonlinearity   {\bf 19},   2695 (2006).

\bibitem{Dmitriev2007} S. V. Dmitriev, P. G. Kevrekidis, N. Yoshikawa, and D. J. Frantzeskakis,  Exact stationary solutions for the translationally invariant discrete nonlinear Schr\"odinger equations,  J. Phys. A.: Math. Theor.  {\bf   40},   1727 (2007).

\bibitem{Susanto2007}
H. Susanto, P. G. Kevrekidis, R. Carretero-Gonz\'alez,  B. A. Malomed, and D. J. Frantzeskakis, Mobility of Discrete Solitons in Quadratically Nonlinear Media, Phys. Rev. Lett. {\bf 99}, 214103 (2007).

\bibitem{AstaMalo}
G. E. Astrakharchik and B. A. Malomed,  Dynamics of one-dimensional quantum droplets, Phys. Rev. A {\bf 98}, 013631 (2018).

\bibitem{Parisi1} L. Parisi,  G. E. Astrakharchik,  and S. Giorgini, Liquid State of One-Dimensional Bose Mixtures: A Quantum Monte Carlo Study, Phys. Rev. Lett. {\bf  122}, 105302 (2019).

\bibitem{Parisi2} L. Parisi and S. Giorgini, Quantum droplets in one-dimensional Bose mixtures: A quantum Monte Carlo study, Phys. Rev. A {\bf  102}, 023318 (2020).

\bibitem{Kartashov2022} Y. V. Kartashov, V. M. Lashkin, M. Modugno, and L. Torner, Spinor-induced instability of kinks, holes and quantum droplets, New J. Phys. {\bf  24}, 073012 (2022).

\bibitem{Louis} P.J.Y. Louis, E.A. Ostrovskaya, C.M. Savage, Yu.S. Kivshar, Bose–Einstein condensates in optical lattices: Band-gap structure and solitons, Phys. Rev. A {\bf 
67}, 013602  (2003).

\bibitem{Peli} D.E. Pelinovsky, A.A. Sukhorukov, Yu.S. Kivshar, Bifurcations and stability of
gap solitons in periodic potentials, Phys. Rev. E {\bf 70}, 036618  (2004).

\bibitem{Yang} J. Yang, \textit{Nonlinear Waves in Integrable and Nonintegrable Systems}, SIAM,
Philadelphia, 2010.

\bibitem{VK} N. G. Vakhitov and A. A. Kolokolov, Stationary solutions of the wave equation in a medium with nonlinearity saturation, Radiophys. Quantum Electron. {\bf  16},  783 (1973) [Izvestiya Vysshikh Uchebnykh Zavedenii, Radiofizika, {\bf   16}, 1020 (1973)].

\bibitem{LiMalomed}  \rev{Y. Li, Z. Luo, Y. Liu, Z. Chen, C. Huang, S.  Fu, H.  Tan, and B.  A. Malomed, Two-dimensional solitons and quantum droplets supported by competing self- and cross-interactions in spin-orbit-coupled condensates, New J. Phys. {\bf   19},  113043  (2017).}


\bibitem{Discrete1} \rev{ F.  Zhao,  Z.  Yan,  X.  Cai, C.  Li,  G.  Chen, H. He,  B.  Liu,  Y. Li, Discrete quantum droplets in one-dimensional optical lattices, Chaos, Solitons and Fractals {\bf 152},  111313 (2021).}

\bibitem{Discrete2} \rev{Z. Zhao, G. Chen, B. Liu, Y. Li, Discrete vortex quantum droplets, Chaos, Solitons and Fractals {\bf 162},  112481 (2022).}

\bibitem{SM} See the Supplementary Material for visualization of dynamics of mobile and immobile flat-top quantum droplets.

\bibitem{Burger1999}
S. Burger, K. Bongs, S. Dettmer, W. Ertmer, and K. Sengstock, A. Sanpera, G. V. Shlyapnikov, and M. Lewenstein, Dark Solitons in Bose-Einstein Condensates, Phys. Rev. Lett. \textbf{83}, 5198 (1999).

\bibitem{Denschlag2000}
J. Denschlag, J. E. Simsarian, D. L. Feder, Charles W. Clark, L. A. Collins, J. Cubizolles, L. Deng, E. W. Hagley, K. Helmerson, W. P. Reinhardt, S. L. Rolston, B. I. Schneider, and W. D. Phillips, Generating Solitons by Phase Engineering of a Bose-Einstein Condensate, Science \textbf{287}, 97 (2000).

\end{thebibliography}
\end{document}